\documentclass[twocolumn,trackchanges]{aastex61}

\usepackage{graphicx,amsmath}
\usepackage{times}

\defcitealias{ugarte-urra2015}{Paper I}

\shorttitle{Modeling coronal response with magnetic flux transport and steady heating}
\shortauthors{Ugarte-Urra, Warren, Upton \& Young}

\begin{document}

\title{Modeling coronal response in decaying active regions\\ with magnetic flux transport and steady heating}

\author[0000-0001-5503-0491]{Ignacio Ugarte-Urra}
\affiliation{Space Science Division, Naval Research Laboratory, Washington, DC 20375, USA}
\author{Harry P. Warren}
\affiliation{Space Science Division, Naval Research Laboratory, Washington, DC 20375, USA}
\author{Lisa A. Upton}
\affiliation{High Altitude Observatory, National Center for Atmospheric Research, 3080 Center Green Dr., Boulder, CO 80301, USA}
\author{Peter R. Young}
\affiliation{College of Science, George Mason University, 4400 University Drive, Fairfax, VA 22030, USA}
\affiliation{NASA Goddard Space Flight Center, Code 671, Greenbelt, MD 20771, USA}
\affiliation{Northumbria University, Newcastle Upon Tyne NE1 8ST, UK}


\begin{abstract}
We present new measurements of the dependence of the Extreme Ultraviolet radiance
on the total magnetic flux in active regions as obtained from the Atmospheric
Imaging Assembly (AIA) and the Helioseismic and Magnetic Imager on board the
Solar Dynamics Observatory (SDO). Using observations of nine active regions tracked along
different stages of evolution, we extend the known radiance - magnetic flux
power-law relationship ($I\propto\Phi^{\alpha}$)
to the AIA 335\,\AA\ passband, and the \ion{Fe}{18} 93.93\,\AA\ spectral line in
the 94 \AA\ passband. We find that the total unsigned magnetic
flux divided by the polarity separation ($\Phi/D$) is a better indicator of
radiance for the \ion{Fe}{18} line with a slope of $\alpha=3.22\pm0.03$. We then
use these results to test our current understanding of magnetic flux evolution
and coronal heating. We use magnetograms from the simulated decay of these active
regions produced by the Advective Flux Transport (AFT) model as boundary conditions
for potential extrapolations of the magnetic field in the corona. We then model
the hydrodynamics of each individual field line with the Enthalpy-based Thermal
Evolution of Loops (EBTEL) model with steady heating scaled as the ratio of the
average field strength and the length ($\bar{B}/L$) and
render the \ion{Fe}{18} and 335\,\AA\ emission. We find that steady heating is
able to partially reproduce the magnitudes and slopes of the EUV radiance -
magnetic flux relationships and discuss how impulsive heating can help reconcile
the discrepancies. This study demonstrates that combined models of magnetic flux
transport, magnetic topology and heating can yield realistic estimates for the
decay of active region radiances with time.

\end{abstract}


\section{Introduction} \label{sec:intro}

The emergence of strong magnetic fields on the solar surface gives rise to active
regions, areas of large radiative output in the X-rays and the Extreme Ultraviolet
(EUV) solar spectrum. Their radiance is extremely well correlated with the total
(unsigned) magnetic flux contained in those fields
\citep[see e.g.][]{gurman1974,schrijver1987,fisher1998,benevolenskaya2002,fludra2002,pevtsov2003,vandriel2003,fludra2008,ugarte-urra2015},
but the nature of the coupling between the magnetic field changes and the
modulation of the plasma response in the atmosphere remains to be understood.

Active regions exhibit radiative changes at a large range of spatial and temporal
scales. As the emission of the plasma in the corona is fundamentally dependent on
its hydrodynamic state (density and temperature), a significant part of the studies
have gone into identifying, characterizing and modeling the fundamental structures
of the atmosphere: coronal loops. Recent reviews \citep{klimchuk2006,reale2014}
summarize the progress made in this area by high resolution imaging, spectroscopy
and hydrodynamic modeling, mostly in 1D. Latest efforts include the development of
magnetohydrodynamic models of magnetic flux tubes in 3D with thermal conduction
and radiation \citep{dahlburg2016,reale2016} that allow for forward modeling of
intensities and direct comparisons with observations.

Studying loops in isolation can be challenging particularly when tracking their
evolution in temperature and across different spectral bands. There is always
potential contamination from other loops along the same line of sight, especially
at $\sim$2\,MK. Many studies have therefore opted to model the complete active
region, or even the full Sun, at any given instant as a set of loops and then
compare the integrated emission to remote sensing images
\citep{schrijver2004,warren2006,warren2007,winebarger2008,lundquist2008a,lundquist2008b,dudik2011}.
These sets of models use 0D or 1D hydrodynamics solutions for each individual
loop in the simulated domain and an ad hoc heating. With the improvement in
performance of high-end computing, these comparisons are now being done against 3D
magnetohydrodynamical (MHD) radiative models, with rigid topology and parametrized
heating \citep{mok2005,mok2016} or forcing at the lower boundary that advects the field
and generates the atmospheric heating through ohmic dissipation
\citep{peter2004,gudiksen2005,peter2006,zacharias2009,hansteen2010,bingert2011,martinez-sykora2011,zacharias2011,testa2012b,bourdin2013,olluri2015}.

These two approaches investigate the heating and atmospheric response on short timescales, minutes
to hours, but there is an equally relevant component of active region heating at longer time
scales.  Active region lifetimes span days to weeks with both total magnetic flux and radiance
going through changes of several orders of magnitude \citep[see review on active region evolution
by][]{vanDriel2015}. After the emergence of an active region, where the region grows in size,
magnetic flux and polarity separation, magnetic fields on the surface are
transported by surface flows, namely differential rotation, meridional circulation and
convection. These processes cause the magnetic elements to spread over increasing areas and
cancel with nearby magnetic elements. In this diffusive evolution part of the observed flux is
lost in the noise. The overall result is, first a rapid increase of the total observed magnetic
flux, followed by a slower decay after the peak, that is accompanied by a similar increase and
then drop in the radiative output \citep[e.g.][]{ugarte-urra2012}. We showed in \citet{ugarte-urra2015}
(hereafter \citetalias{ugarte-urra2015}) that a state-of-the-art magnetic flux transport model
such as the Advective Flux Transport (AFT) \citep{upton2014,upton2014b} model can make realistic
predictions of the total magnetic flux decline in an active region. Different coronal heating
theories predict different scalings for how the change in the magnetic field can affect the
atmosphere \citep[e.g.][]{mandrini2000}. One scaling that has been most successful to reproduce
observables in the corona is heating that depends on the field strength and length of
coronal loops \citep[e.g.][]{warren2006,warren2007,lundquist2008b}, two quantities that change
over the evolutionary time scales of active regions.

In the present paper, we investigate whether we can make use of this improved
understanding of magnetic flux transport to model the coronal response over long
periods of time. We use the AFT simulated magnetic field evolution of nine active
regions as a boundary condition for a quasi-static steady heating model, a steady
heating scenario that has proven to be successful to model
the high temperature plasma in active region snapshots. We show that we can indeed
predict the decline of EUV emission as a function of time and
magnetic flux, and we discuss potential improvements when incorporating more
sophisticated models. Furthermore, we find that the observed \ion{Fe}{18} emission
is better described by the combination of the magnetic flux and the mean loop length
than by the magnetic flux alone. This provided further evidence for a coronal
heating rate that is parameterized by magnetic flux and loop length.

\section{Observations} \label{sec:obs}

We looked at a set of nine active regions previously selected for a study of their long-term evolution
in the context of magnetic flux transport and decay \citepalias[see][]{ugarte-urra2015}. The regions, from the
period January 2011 -- July 2013, were chosen for their isolation, i.e. regions minimally impacted by
neighboring active regions. In that investigation we used integrated 304 \AA\ light curves as a proxy for
active region development (emergence, growth and decay) and found that it can also be used as a proxy
for total unsigned magnetic flux. Several regions were observed from emergence (11158,11672,11726,11765)
while others are seen in evolved stages. The 304\,\AA\ evolution lightcurves can be consulted in Figure
3 of \citetalias{ugarte-urra2015}.

In the current study, we turn our focus to the response of those regions in the corona.  We
inspected EUV coronal images from the Atmospheric Imaging Assembly (AIA) \citep{lemen2012} on board
the {\it Solar Dynamics Observatory} \citep{pesnell2012}. AIA takes high cadence (12\,s) and
high-resolution (0.6\arcsec\ per pixel) images of the Sun in ten narrow-band filters.
For our study, active regions were tracked from limb to limb at a cadence of 1 hr,
sufficient to characterize the gradual evolution. The
94\,\AA\ band is centered at the spectral line \ion{Fe}{18} 93.93\,\AA, with a formation
temperature of 7$\times10^6$\,K. This line dominates the emission in active region and flare
conditions \citep{odwyer2010,testa2012,teriaca2012}, but contributions from \ion{Fe}{8},
\ion{Fe}{10} and \ion{Fe}{14} ions can be important in the quiet Sun.  As shown by
\citet{warren2012} and \citet{teriaca2012}, the ``hot'' \ion{Fe}{18} line emission in the channel
can be isolated by removing the contaminating ``warm'' component from a weighted combination of
emission from the 193\,\AA\ and the 171\,\AA\ channels. In \citet{ugarte-urra2014} we used this
method to study the temporal evolution of individual coronal loops in \ion{Fe}{18} emission. In
that study we argued that the \ion{Fe}{18} line is a good diagnostic for the study of loops in
active region cores because the temperature response contains the peak of their emission
measure distribution ($\sim4\times10^6$\,K) and it is narrower than X-ray broad-band filters observing in those
temperature ranges. In the present paper, we are interested in the global properties of the active
region and we integrate the \ion{Fe}{18} counts to derive the total radiative output in that line. We
also use images in the 335\,\AA\ band, dominated by \ion{Fe}{16} with a formation temperature of
2.8$\times10^6$\,K. The latter is a more traditional temperature range for the study of loops and
active regions and facilitates the comparison to other results. We applied a time dependent
sensitivity correction for 335\,\AA\ images that is available in the standard software distribution
({\tt aia\_get\_response.pro}). In the rest of the paper we refer to counts per second in the 335 channel,
and this always means the counts per second corrected to be as if the images were obtained at the start of
the mission rather than the actual measured counts. In contrast to the 94\,\AA\ passband, we do not have an
empirical method to isolate the \ion{Fe}{16} from other cooler contributions \citep{odwyer2010} and we perform
the analysis over the whole filter spectrally integrated signal.

We computed the total unsigned magnetic flux for every active region as
a function of time: $\Phi(t) = \int |B_z(t)| dA$ where $A$ is area on the surface
and $B_z$ is the vertical magnetic field. We used line-of-sight magnetograms from
the Helioseismic and Magnetic Imager/SDO \citep{Scherrer2012,Schou2012}, correcting
for line of sight projection angle effects for both the pixel areas and the observed
flux densities.

\section{Magnetic flux and EUV radiance}

It is well known that the EUV and X-ray response of the solar corona scales
up with the magnetic flux content. This is normally presented in the form of a
magnetic flux-luminosity relationship that has been explored for a variety of solar phenomena
such as active regions, coronal bright points, quiet Sun and even other stars
\citep{pevtsov2003}. In the case of active regions, these scatter plot relationships
have been generally constructed statistically, where different magnetic fluxes and
luminosities (or radiances) are provided by active regions of different size and
strength \citep{gurman1974,schrijver1987,fisher1998,fludra2002,fludra2008}. There are also examples of scatter plots made out of the evolving properties
of a single region \citep[e.g.][]{vandriel2003} or a coronal bright point \citep{ugarte-urra2004}.
In this study, we combine both approaches. We first track the total unsigned flux and EUV
radiance changes in single active regions from birth to decay. Then we combine the results
for multiple cases, the nine active regions in our dataset. In \citetalias{ugarte-urra2015} we
did exactly that for the 304 \AA\ intensities. Here we extend it to the 335 \AA\ channel
and the \ion{Fe}{18} component in the 94 \AA\ channel.

\begin{figure*}
\plottwo{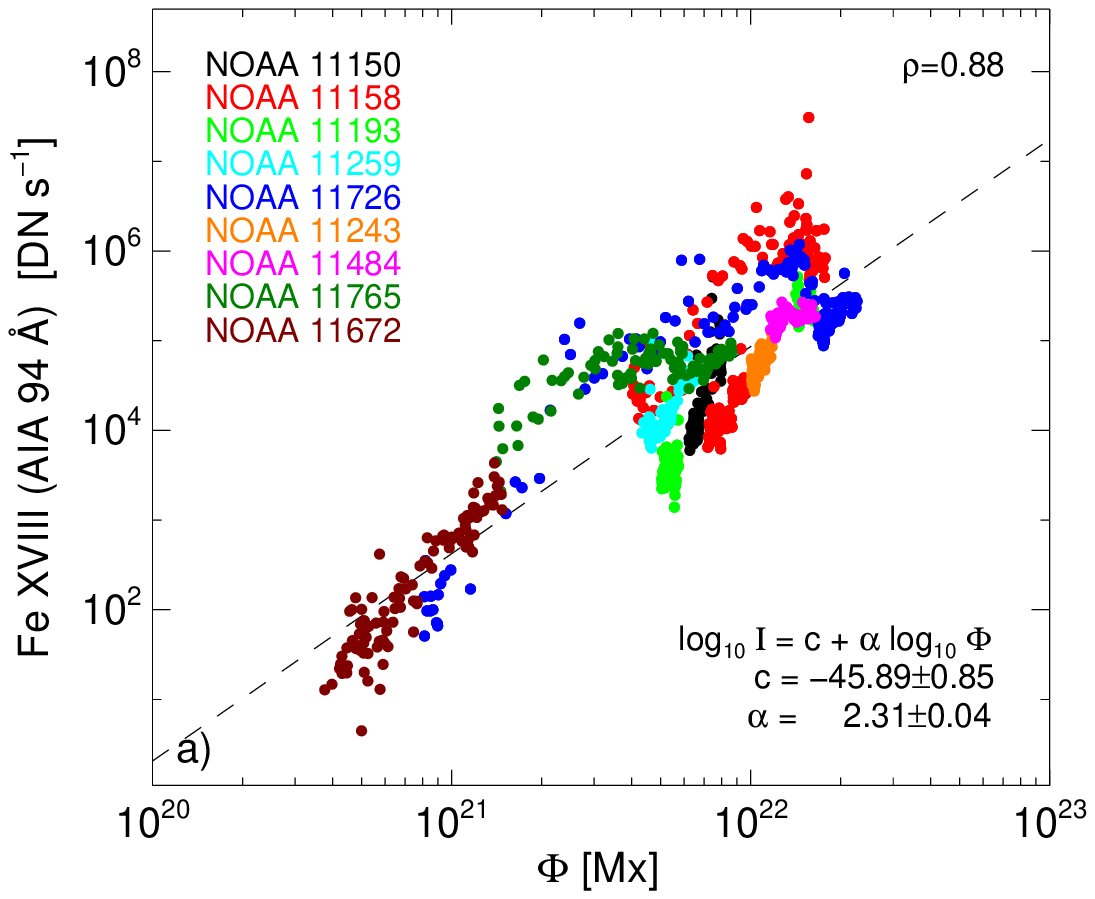}{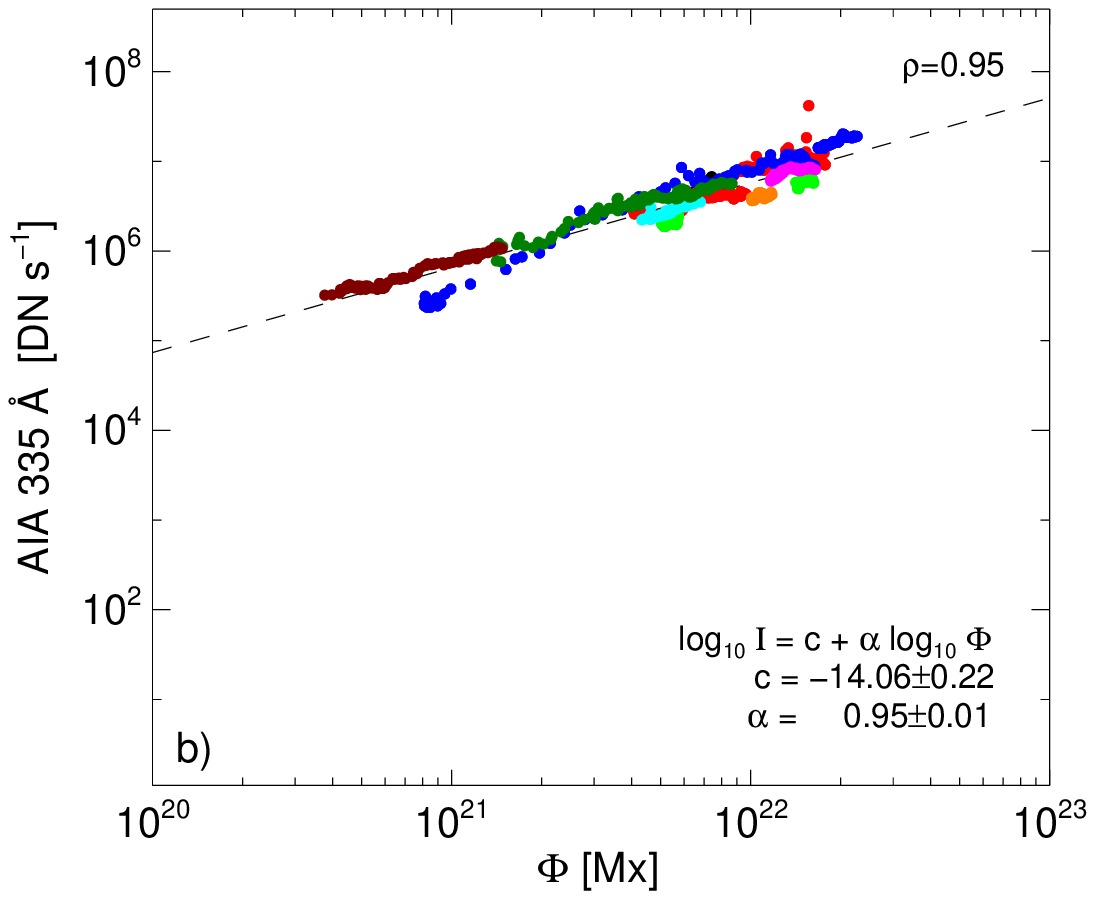}
\plottwo{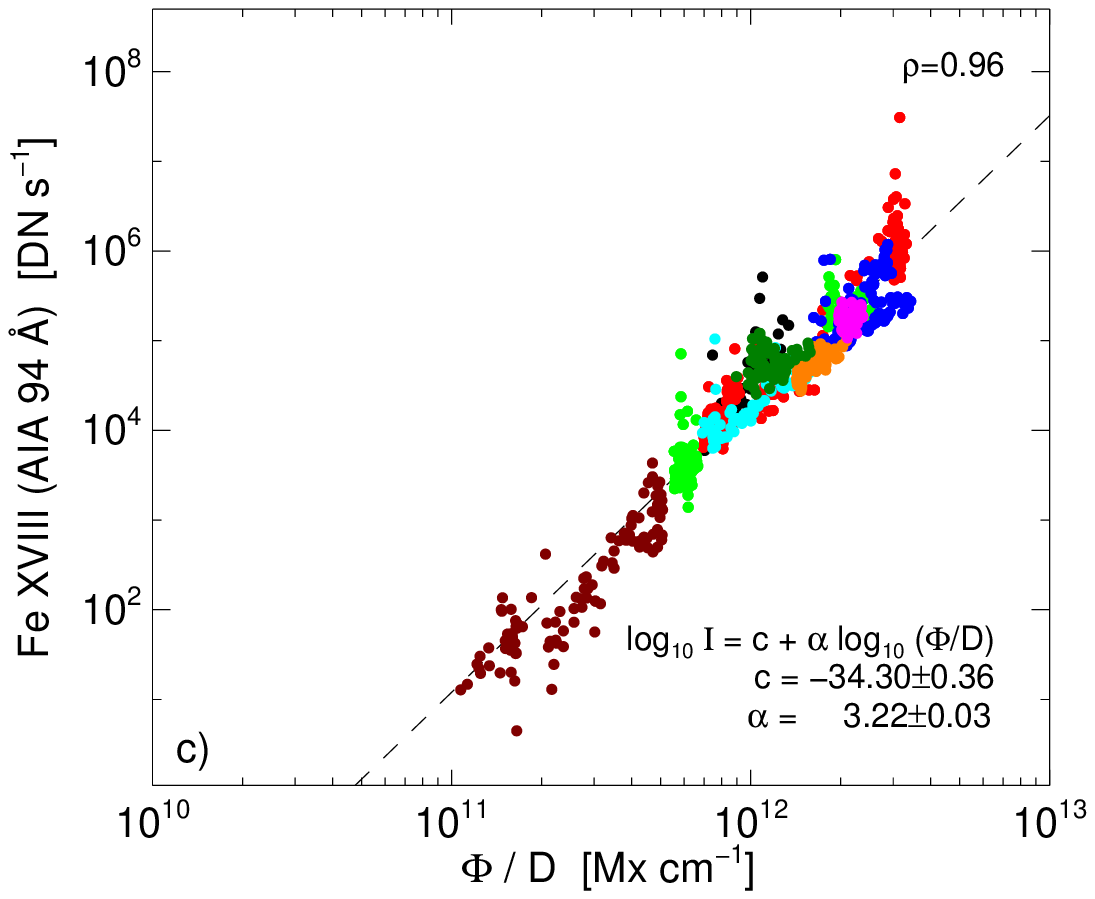}{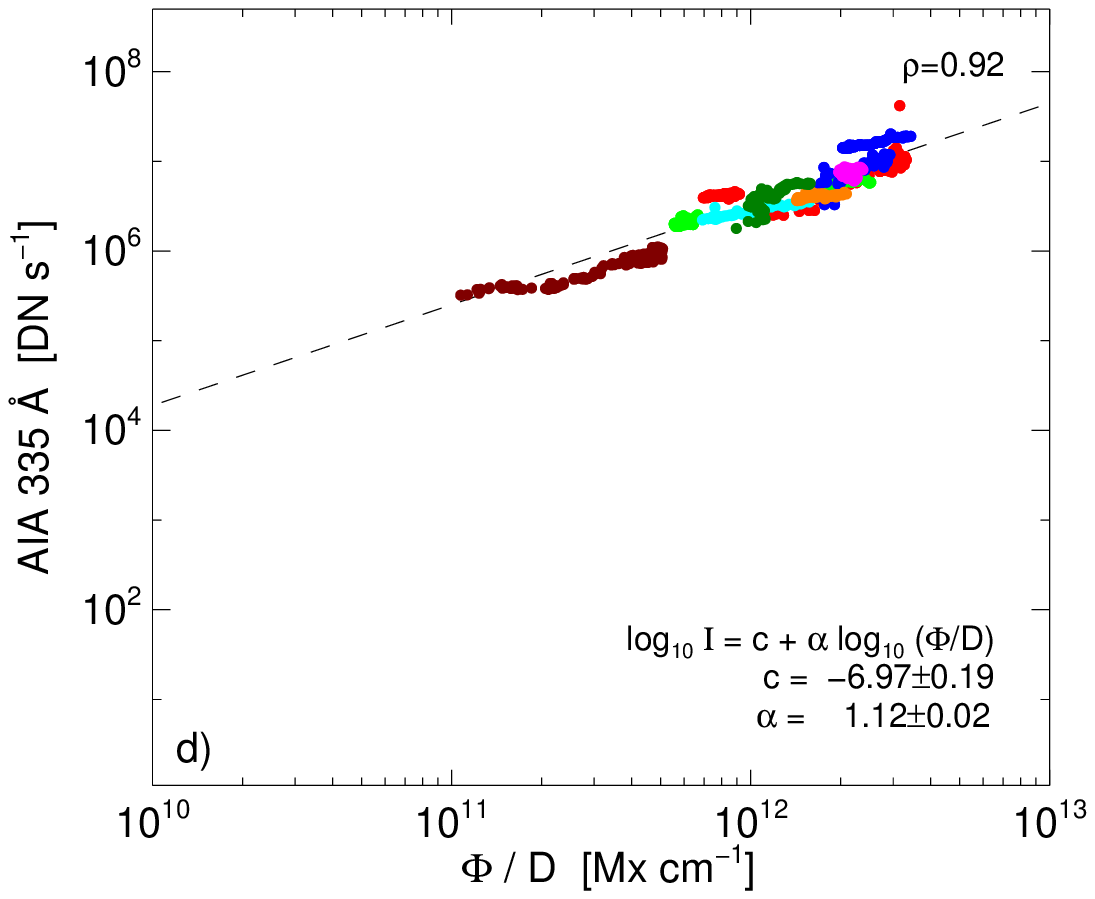}
\caption{Scatterplot of the EUV radiance and total unsigned magnetic flux for nine
active regions. The different points in each region correspond to different times
in the evolution. Panels c) and d) show the total unsigned magnetic flux divided by
the separation between the centroid of the two polarities. Dashed  lines
are linear fits to the logarithmic quantities. $\rho$ is the Pearson correlation coefficient.
\label{fig:F-L1}}
\end{figure*}

\begin{figure*}
\includegraphics[width=\textwidth]{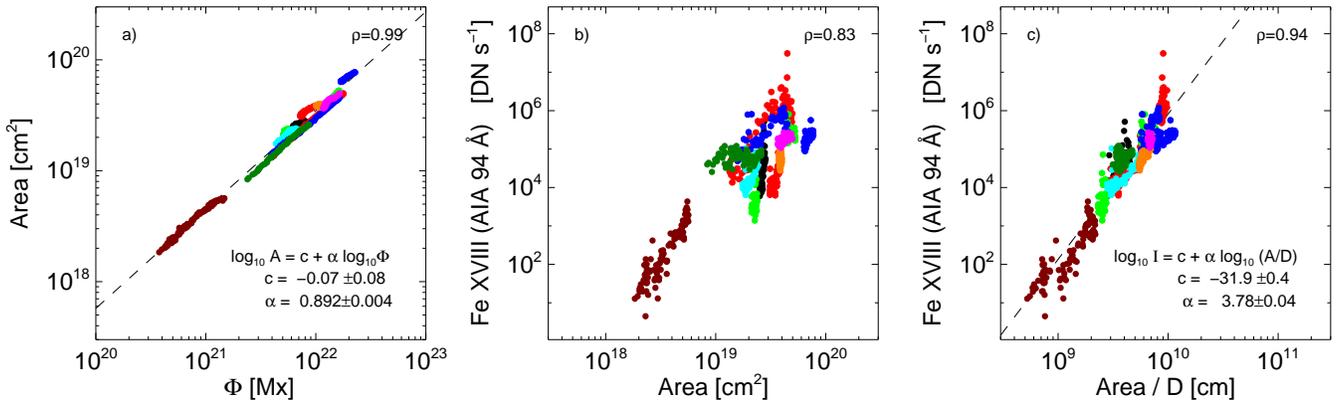}
\caption{Scatterplot of the a) active region magnetic area versus total unsigned
magnetic flux, b) \ion{Fe}{18} radiance versus magnetic area, and c) \ion{Fe}{18} radiance
versus the ratio of magnetic area to polarity separation. Dashed lines are linear fits.
$\rho$ is the Pearson correlation coefficient. Colors represent different
active regions, same as Figure \ref{fig:F-L1}.  \label{fig:area}}
\end{figure*}

Figure \ref{fig:F-L1} shows the scatter plot of the total number of AIA \ion{Fe}{18}
and 335 \AA\ counts per second as a function of total unsigned magnetic flux for
each active region in the dataset. Each datapoint corresponds to a particular
instant in the evolution of the region while it transits within 60$^{\circ}$ of Sun center.
In the case of NOAA 11158, 11193 and 11726, the longest lived regions in the dataset
\citepalias[see Figure 1 in][]{ugarte-urra2015}, we include data points for two
rotations. Counts were integrated within a rectangular field-of-view with sides
ranging from 100\arcsec\ to 325\arcsec\ depending on the size of the active region and
above a threshold of 2 and 4 DN\,s$^{-1}$ for \ion{Fe}{18} and 335 \AA\ images respectively.
Following \citet{fludra2008} we only considered magnetic flux densities in the range
90 -- 900 G, within the same rectangular field-of-views. The 900 G is a conservative limit to exclude sunspots. Preliminary
calculations with a lower threshold of 20 G and including sunspot flux, as used for
304 \AA\ in \citetalias{ugarte-urra2015}, resulted in similar overall results with
minor changes ($\lesssim0.1$) in the slopes.

The log-log plots show that, as expected, the correlation between the two quantities for both
channels is high and is well represented by a power law fit ($I \propto\Phi^\alpha$).  The slope in
335\,\AA\ is $\alpha = 0.95$ close to the 1.1 -- 1.3 values obtained for \ion{Fe}{16} from CDS
spectroscopic measurements \citep{fludra2002,fludra2008}, although it is hard to make direct comparisons
between these studies when different units are used as \citet{fludra2008} rightly pointed out.
 The slope for \ion{Fe}{18} is much
steeper (2.31), which is a consistent trend with previous estimates (1.6 -- 2) of hotter plasma
using X-ray broad band filters \citep[e.g.][]{warren2006,vandriel2003,benevolenskaya2002}, but
\citet{fisher1998} reported 1.19 in luminosity units. The dispersion for the \ion{Fe}{18} slope is
larger than 335 with trends for the individual regions that seem to diverge at times from the overall
power law fit.  Flaring activity has not been filtered out in the analysis and could contribute to the
dispersion, but most regions only produced C-class flares and only NOAA 11158 produced X and M-class
flares, visible as the extreme values in \ion{Fe}{18} counts in Figure \ref{fig:F-L1}a.

Previous studies already conclude that the area of the active region, highly correlated
to the total magnetic flux (Figure \ref{fig:area}a), is the dominant factor in the flux
relationship to the radiance \citep{schrijver1987,fisher1998,fludra2008}. This is evident
when we see the similitude between the \ion{Fe}{18} radiance dependence with the total magnetic area
(Figure \ref{fig:area}b) and the flux dependence in Figure \ref{fig:F-L1}a.

In terms of the coupling between magnetic flux and radiance, several studies identify the ratio of
the loop's average field strength to its length ($\bar{B}/L$) as a volumetric heating
scaling that has been successful to model high temperature radiation in active regions
\citep{warren2006,warren2007,lundquist2008b}.
If the heating of loops depends on $\bar{B}$ and $L$, it seems natural to consider
how important for the radiance is the spread of the corresponding total magnetic
flux and area. A similar argument was made by \citet{fludra2003} who tested it with
CDS \ion{Fe}{16} data and found a very weak dependence on an average loop length.

In the Figures \ref{fig:F-L1}c and \ref{fig:F-L1}d we show how the relationship changes
when we add a characteristic length $D$ to our dataset.
We have chosen $D$ as the separation between the magnetic flux density weighted
centroids of the two polarities (within 90 -- 900 G), calculated as the great-circle distance
\begin{eqnarray}
D & = & R_{\sun} \arccos (\cos\theta_1 \cos\theta_2 \cos(\phi_1-\phi_2) + \sin\theta_1 \sin\theta_2)
\nonumber
\end{eqnarray}
where $\theta$ and $\phi$ are the latitude and longitude of the centroids and
$R_{\sun}$ the solar radius. These scatter plots have fewer points per active
region because we had to visually filter out instances in the evolution, e.g.
emergence, when the automatic distance measurements failed.
It is worth noting that despite the already high
correlation between the \ion{Fe}{18} counts and the magnetic flux, adding the
characteristic length does improve the correlation and the power-law fit, and
increases the slope to 3.22 suggesting that the total flux per characteristic
separation length is a better indicator for hot lines such as \ion{Fe}{18}. This
is also true for the area (Figure \ref{fig:area}c) meaning that in terms of
intensity it is relevant how spread that area is, not surprising if the heating
is inversely proportional to the loop length. Until now, however, this
dependance on length has remained undetected. One possible explanation is the
spectral line selection. We have argued that \ion{Fe}{18} is a particularly good
diagnostic in active regions for its formation temperature and its spectral purity.
Most studies use cooler EUV lines (e.g. \ion{Fe}{16}), or X-ray broad band filters
that integrate signal from those lower temperatures. Indeed, the influence of the
separation for 335 \AA\ is minimal (even producing a slightly worse fit),
which is consistent with previous results such as \citet{fludra2003} that report a weak
dependence for \ion{Fe}{16}. In this line, the filling of the active region coronal
volume with signal is larger than that of \ion{Fe}{18}, with more loops at
different lengths contributing to the integrated intensity.
In a later discussion we also show that \ion{Fe}{18} and \ion{Fe}{16} lines can be
diagnostics for different phases in the evolution of a coronal loop.

In the following section we use global active region modeling to see whether
these measurements are consistent with our current understanding of heating
in active regions.

\begin{figure*}
\centering
\includegraphics[angle = 90,width=18cm]{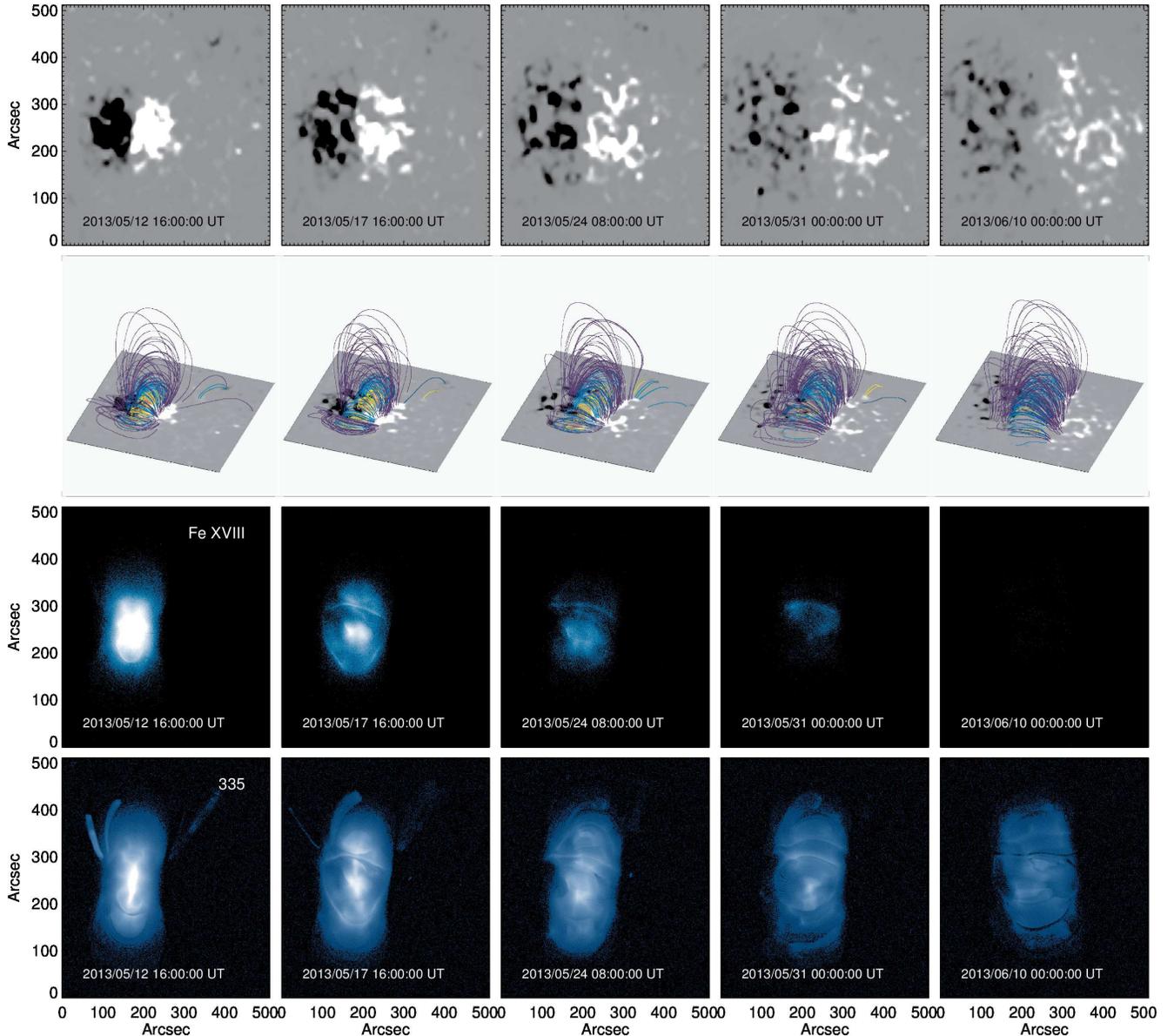}
\caption{Representative simulation of active region decay that combines three models:
magnetic flux transport, magnetic topology and hydrodynamic heating. First row: vertical magnetic
field evolution as simulated by the Advective Flux Transport model. Second row:
potential magnetic field extrapolation of the AFT magnetograms. Third and fourth
rows: integrated predicted radiance in \ion{Fe}{18} and 335 \AA\ for a hydrodynamic
state where each extrapolated loop is heated steadily (equation~\ref{eq:1}).
This example corresponds to the magnetic flux evolution of NOAA 11726.
\label{fig:ar11726}}
\end{figure*}

\section{Modeling} \label{sec:model}

The flux-radiance relationships tell us that by knowing the total magnetic flux
of an active region, we are able to estimate the radiance of that region in X-rays
and the EUV. In \citetalias{ugarte-urra2015} we argued that a state-of-the-art magnetic
flux transport model such as the Advective Flux Transport (AFT)
\citep{upton2014,upton2014b} model is already capable of making realistic predictions
of the total magnetic flux decay in an active region. Therefore relationships
such as those in Figures~\ref{fig:F-L1} combined with the AFT
model flux predictions, in principle, allow  us to make reasonable estimates of
AIA \ion{Fe}{18} and 335 \AA\ counts.

In the following sections, we go beyond the empirical description and investigate
whether the model of the flux decay combined with a simple coronal
heating model can reproduce these observed relationships. The actual coronal
heating mechanism is still not known and while there are some 3D MHD efforts
to  study the problem from first principles, simulations are still constrained to brief
periods of time of individual loops or small active regions. As we are looking at
a set of nine active regions evolving over periods of weeks and months, our approach
is simpler. We look at snapshots of the active region magnetic evolution determined by the
AFT model and we model the 3D magnetic topology with a potential magnetic field
extrapolation. The coronal response of each coronal loop in the volume is then
simulated with a 0D hydrodynamic model in an ad hoc steady heating approximation.
While steady heating may be difficult to reconcile with the evolution of loops
observed at $\lesssim2\times10^6$\,K \citep[but see][]{mikic2013}, it can help us
understand the role that the magnitudes used in the parametrization of the heating
play in the overall energy budget.

\subsection{Magnetic flux decay}
We model the magnetic flux decay of the active regions in our dataset using
the AFT model, a surface flux transport (SFT) model. SFT models simulate the
displacement of the magnetic flux on the surface through the following flows:
differential rotation, meridional circulation, and the cellular and turbulent motions of
convection. AFT is distinct from other SFT models in the way it treats the transport
of convective motions. Most models parametrize that transport as a diffusive process.
AFT models it explicitly using a convective velocity field with the spectral
characteristics of the flows observed on the Sun \citep{hathaway2010}.
A detailed description of the model can be found in \citet{upton2014,upton2014b}.
\citetalias{ugarte-urra2015} describes its application to our current active region dataset.

The baseline of the model assimilates magnetic flux density measurements from
line-of-sight magnetograms, only available on the Earth side view of the Sun. The
model evolves the flux everywhere else, where data is not available, until the region
rotates back into view and the assimilation process resumes. To study the performance
of the model for timescales longer than the fourteen day period that the active region
rotates across the back of the Sun, assimilation can
be turned off at any moment thus letting the flows govern the flux evolution for
the remaining time of the experiment. In \citetalias{ugarte-urra2015} we used this
option to forecast the decay of all active regions in the dataset
from the moment they reached a peak in the total unsigned magnetic flux until full decay.
In three of the regions (NOAA 11272, 11484,
and 11726) that peak in flux happened while on the back side, so we used the 304 \AA\
proxy to insert a bipolar region in the AFT maps and then allowed them to decay. The study
showed that the AFT description provides realistic predictions of the decaying
flux for periods longer than a solar rotation.

In the current study, we use the actual AFT magnetograms from those forecast
calculations as the boundary conditions to model the magnetic field distribution
in the corona. The AFT maps are Carrington projections of the full Sun at 0.35$^\circ$
resolution. We selected 9 -- 11 AFT magnetograms per active region, encompassing the
complete region's lifetime from peak magnetic flux to the time when the 304 \AA\ light
curve had decayed to background levels. We deprojected the maps to heliographic coordinates at a spatial resolution
of 1\arcsec\ and extracted subareas of 512$\times$512 pixels around the center of
the active regions. The top row of Figure~\ref{fig:ar11726} shows  a subset of the AFT
simulated magnetograms for the NOAA 11726 fluxes. The simulated surface field
distribution has many of the ingredients of an active region evolution besides a
realistic flux decay: the polarities diffuse away and
the surface flow pattern creates a reticulated field distribution characteristic
of the active region plage.

\subsection{Magnetic topology}
Given the frozen-in conditions in the solar corona and assuming field aligned plasma flows, one way
of modeling a full active region is by treating the hydrodynamics of each
magnetic flux tube or loop independently. In this approximation, to identify the loops we
need a description of how the magnetic field is distributed in the atmosphere. We use the minimum
energy state, the potential field approximation --- that is we extrapolate the field from the
magnetograms into the corona using Fourier transform solutions to the Laplace's equation for the
scalar potential in a Cartesian coordinate system \citep[e.g][]{nakagawa1972}.  This current free
approximation is not necessarily the most accurate description of the magnetic topology in an
active region. Active regions are known to store magnetic energy in the form of currents, so
present day state-of-the-art extrapolations are obtained in the non linear force free (NLFF) approximation
\citep{wiegelmann2012}.  We find, however, justified to use a simpler topological model in our
study for two reasons.  First, one of the interesting implications of the flux-luminosity
relationship is that the radiative output correlates well with total unsigned magnetic flux, a
quantity that is largely independent of the 3D topology. \citet{fisher1998}, for example,
  failed to find any correlation between X-ray luminosity and the non-potential components of the
  field inferred from vector data.
Two active regions with the same magnetic flux, but different topologies and potentiality, should
have approximately the same EUV emission, ignoring transient events. Second, we are not trying to
reproduce the exact morphology and energetics in these regions. Instead, our goal is to find out
whether a model that includes our basic understanding of flux transport, topology and heating is
capable of reproducing observables in the active region decay. More complex descriptions will
follow.

We used the AFT line-of-sight magnetograms as boundary conditions for the
extrapolation and a field line tracing algorithm to identify field lines for
every pixel (1\arcsec\ resolution). We limited the search to magnetic flux
densities in the range 20 -- 900 Mx cm$^{-2}$ avoiding low signal-to-noise
levels and the core of sunspots as they do not generally emit in the EUV. The
panels in the second row of Figure~\ref{fig:ar11726} are examples of the
magnetic topology model for NOAA 11726.

\subsection{Coronal model}

To simulate the total radiance of the active regions in the \ion{Fe}{18} and 335 \AA\ passbands, we
model the active regions as a set of independent loops. In these optically thin emission bands the
loops' intensities are fundamentally dependent on the plasma electron density and temperature at
any given instant, with the total active region intensity simply the sum of all the loops. For
every region we assign one loop per field line in the extrapolation and model their hydrodynamics
using a 0D coronal model called ``Enthalpy-based Thermal Evolution of Loops'' EBTEL
\citep{klimchuk2008,cargill2012} that has been shown to provide results comparable to a
state-of-the-art 1D model. We have used it in the past to study the envelope intensity lightcurve
of a set of loops \citep{ugarte-urra2014}.  These coronal models, with their restricted field
aligned dynamics, prevent us from addressing cross-field effects that can be important to
understand the nature of the heating \citep[e.g][]{dahlburg2016,reale2016}, but they allow us to do
calculations that are not manageable today by 3D MHD codes.

\begin{figure*}
\centering
\includegraphics[width=\textwidth]{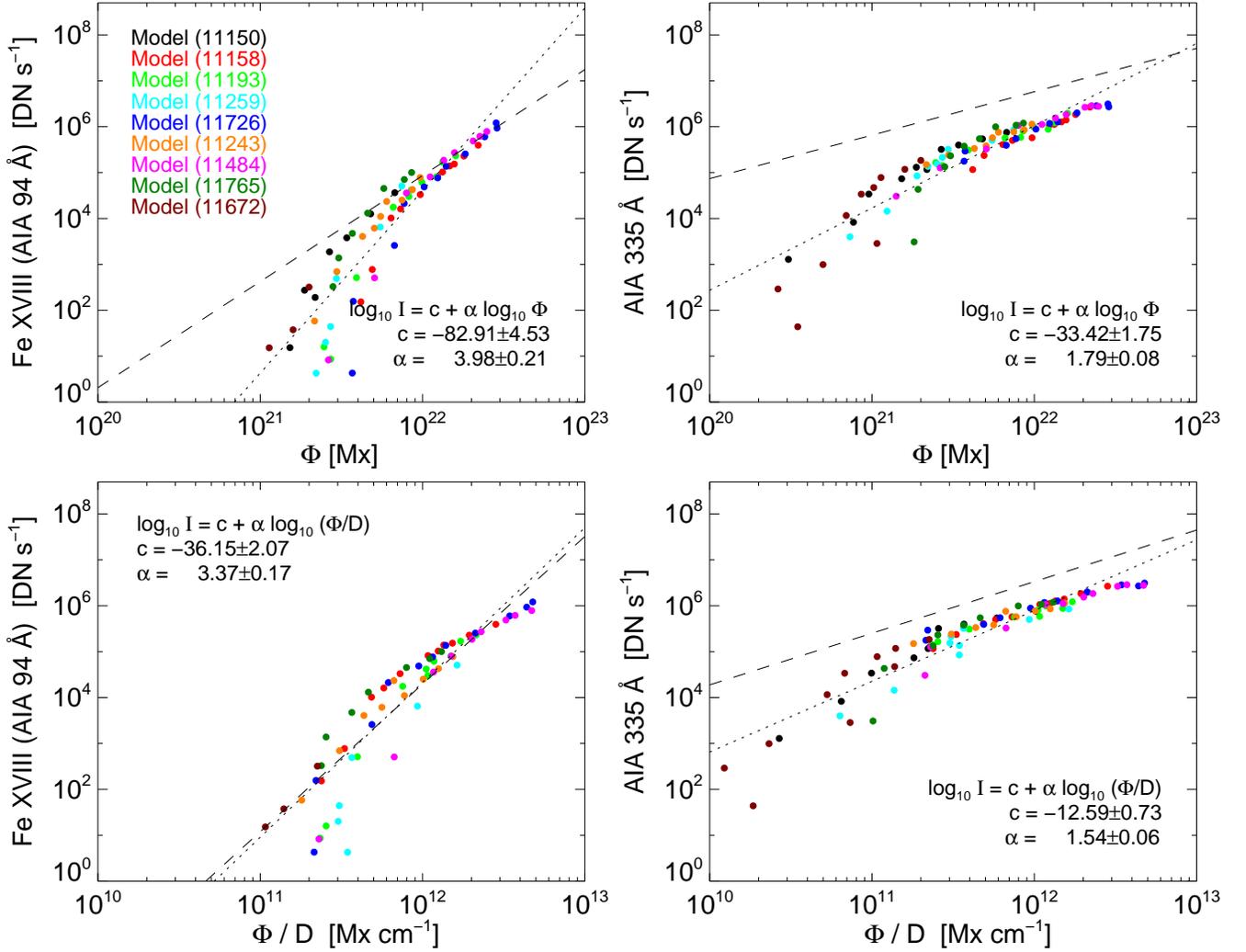}
\caption{Scatterplot of the EUV radiance and total unsigned magnetic flux for the simulated active regions,
not including any background intensities. Dotted lines are linear fits to simulation
data. Dashed lines are the fits to the observational data in Figure~\ref{fig:F-L1}.
\label{fig:simF-L}}
\end{figure*}

EBTEL computes the evolution in time of spatially averaged properties of the loop
such as coronal density and temperature in response to a variable heating. The only inputs
are the loop's length and the volumetric heating rate. The model was conceived
to study the parameter space of impulsive heating events, but in the context of this
study we have chosen to drop the time dependency and investigate a steady heating
scenario with a volumetric heating rate of the form
\begin{eqnarray}\label{eq:1}
\epsilon_H = \epsilon_0 {\bar{B} \over \bar{B_0}} {L_0 \over L}
\end{eqnarray}
where $L$ is the total length of the loop and $\bar{B}$ is the average field
strength. \citet{warren2006} found that this functional form for the heating
produces emission at high temperatures that is consistent with observations.
In that study they choose $\epsilon_0$ so that the loop $\bar{B}=\bar{B_0}=76$ G
and $L=L_0=29$ Mm has an apex temperature of 4 MK and use a filling factor
to match the intensities. Here we choose the same values for $L_0$ and $\bar{B_0}$
but leave $\epsilon_0$ as a free variable to match the observed intensities.
In the context of other loop studies,
the heating is coronal, steady and uniformly distributed and leads to a time
independent  equilibrium that balances heating with radiative losses.

For each loop, we compute the \ion{Fe}{18} and 335 \AA\ counts using the AIA instrument response
functions as a function of temperature, the coronal temperature and density returned by EBTEL
given the $\bar{B}/L$ heating rate, and the loop's volume. We do not include the differential
emission measure from the transition region.
To estimate the AIA \ion{Fe}{18} response with no cool contribution, we compute
the emissivity of the spectral line as a function of temperature with CHIANTI and scale it down
to match the high temperature peak in the 94\,\AA\ response.
We assume all loops to be cylinders of radius 350 km and length $L$ determined by
the extrapolation.  The radius choice is convenient given the 1\arcsec\ (725 km) resolution of the
magnetograms and the extrapolation of one field line per pixel. The assumption is also justified
based on a recent studies of loop size measurements that argue that many observed loops are single
monolithic structures at the scales of hundred of kilometers and are resolved by current
instrumentation \citep{brooks2012,brooks2013,aschwanden2017}. A constant loop width assumption
is consistent with observations \citep{klimchuk2000,lopezfuentes2006}, but still unexplained by
magnetic field models. In the approximately force-free corona, these models predict a flux
tube expansion with height result of the magnetic field strength falloff.

The total counts for each loop is then distributed along the corresponding 1\arcsec\
pixels of the extrapolation volume and projected to the plane of the image taking
into account the area factor conversion from the default 0.6\arcsec\ per pixel in
the AIA response. To mimic the optical response of the AIA instrument to a 350 km
radius source, the projected  images are convolved with a Gaussian filter with a
full width half maximum (FWHM) of 1.47 pixels (1056 km), the apparent size for
such a source in an instrument with FWHM point spread function of 1.14\arcsec,
that of AIA \citep{grigis2012}. As a final step, we add noise to the images. For
the 335 \AA\ filter images we apply a poissonian photon distribution to the
uncertainty estimates that can be obtained from the standard AIA software distribution
({\tt aia\_bp\_estimate\_error.pro}) in photons. In the case of \ion{Fe}{18}, there
are no estimates because our images are a processed version of the 94 \AA\ filter
images, and we use an empirical relationship where the uncertainties scale as
(DN\,s$^{-1}$)$^{1/3}$, obtained in \citet{ugarte-urra2014} from the standard deviation
of sets of five consecutive 12s cadence images. Examples of simulated images
can be seen in Figure~\ref{fig:ar11726} for NOAA 11726. While the
magnetic flux decay is similar to that of the observed region, the
magnetic topology and the simulated morphology in the EUV are not expected to
match the observations.

Figure~\ref{fig:simF-L} presents the model results based on all assumptions above.
The panels show the integrated counts per second as a function of total magnetic
flux and flux divided by polarities separation for both \ion{Fe}{18} and 335 \AA\
for all regions. As with the observations, we only considered pixels with
radiance values above 2 DN s$^{-1}$ (\ion{Fe}{18}) and 4 DN s$^{-1}$ (335 \AA),
and magnetic flux densities within 90 -- 900 G. The dashed lines in Figure~\ref{fig:simF-L}
are the linear fits to the AIA and HMI data from Figure~\ref{fig:F-L1} for reference.
To find this match, we only varied $\epsilon_0$.
The adopted value is 0.005 ergs cm$^{-3}$ s$^{-1}$ which
brings the \ion{Fe}{18} at high counts in close agreement to the measurements. Increasing
$\epsilon_0$ to 0.010 results in an order of magnitude increase for the \ion{Fe}{18}
radiance and just about a factor of 2 for 335 \AA, supporting our earlier argument
that the \ion{Fe}{18} line is a better heating diagnostic. Given the simplicity of
the assumptions, it seems remarkable how close the simulated flux - radiance pairs
are to the actual measurements in terms of absolute numbers and trends (slopes).
Recall that the only data constraints to the model are the peak magnetic
flux obtained from the 304 \AA\ proxy that starts the AFT model forecast and the
choice of $\epsilon_0$ in the heating to match the \ion{Fe}{18} radiance levels.
There are, nevertheless, obvious discrepancies that suggest that the model is still
far from being a complete satisfactory description of the observations. We discuss
below those discrepancies and potential improvements.

\begin{figure*}
\centering
\includegraphics[width=\textwidth]{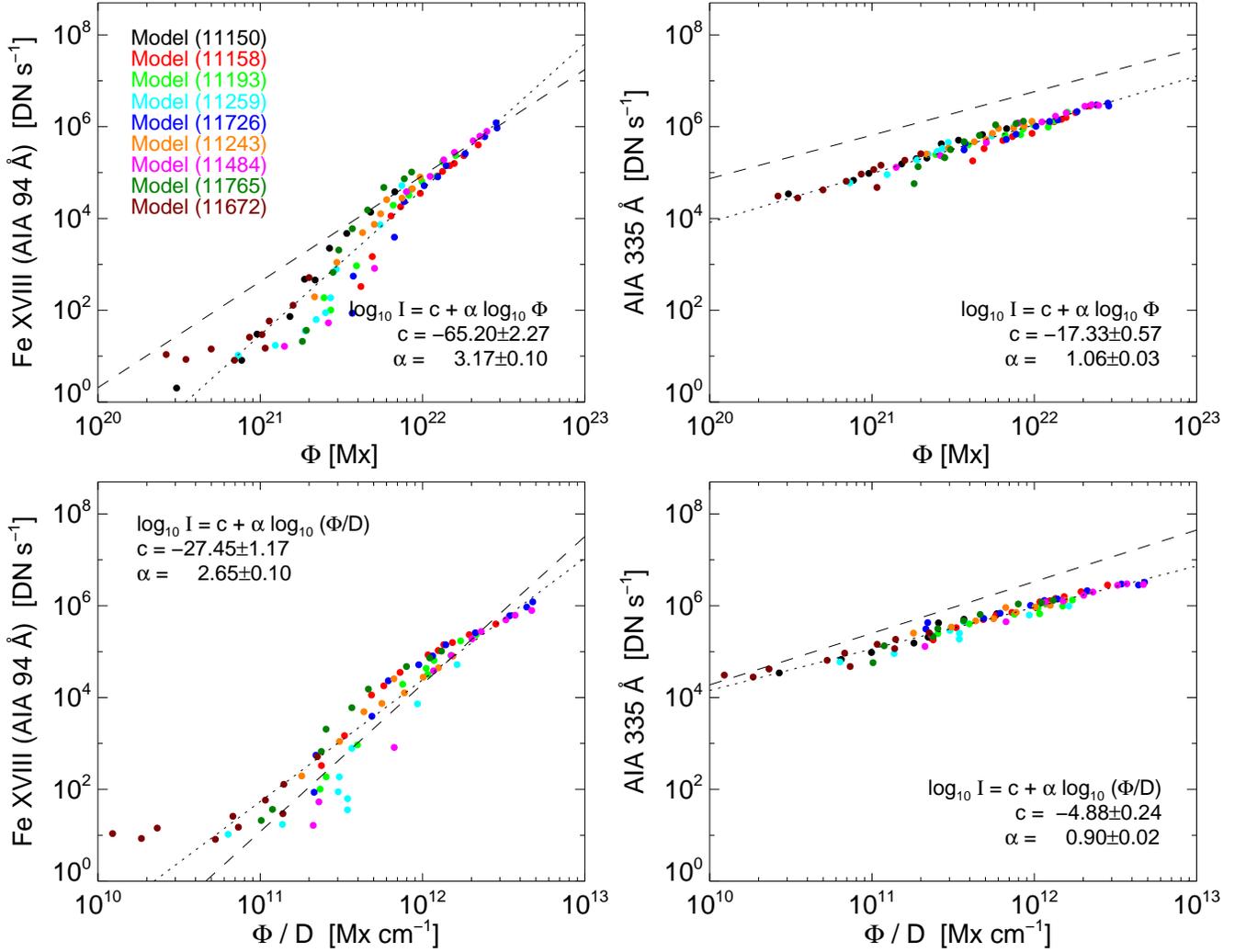}
\caption{Scatterplot of the EUV radiance and total unsigned magnetic flux for the
simulated active regions, including a background intensity of 0.1 and 1.5 DN\,s$^{-1}$
for \ion{Fe}{18} and 335 \AA\ respectively. Dotted lines are linear fits to simulation
data. Dashed lines are the fits to the observational data in Figure~\ref{fig:F-L1}.
\label{fig:simF-Lbackg}}
\end{figure*}

\subsection{Discussion}

We have shown in the previous section that the model is able to explain the order
of magnitude change in both radiance and magnetic flux in the evolution of active
regions. This is consistent with the results from \citet{warren2006} that used
hydrostatic solutions with a $\bar{B}/L$ scaling to reproduce the X-ray -- flux
relationship using real magnetograms. Our model with simulated magnetograms is
able to reproduce the \ion{Fe}{18} slope for $\Phi/D$, and for $\Phi$ at high
radiance values, but a non-linear trend towards low fluxes seems apparent for both.
This is also noticeable in 335 \AA. In \ion{Fe}{18}, as the region decays and
disappears, radiance reaches the few hundred DN~s$^{-1}$ level from the added
contribution of several pixels close to the threshold level, so we may be reaching the
limit of what can be diagnosed with a line like \ion{Fe}{18} and an AIA instrument
response. While in \citetalias{ugarte-urra2015} we showed that the AFT model is able
to predict the total flux of select active regions to within a factor of 2, there
have not been detailed statistical comparisons between the AFT simulated active
region evolution and observations. Potential discrepancies between the magnetic
field  distribution in AFT and real magnetograms could propagate into the heating
scaling and need to be investigated.

The main discrepancies for the 335 \AA\ scatterplot are total counts, with a
systematic underestimation in the model, and a drop of radiance at small fluxes.
The drop at small fluxes can be explained if we consider a background intensity.
The model does not include loops that are not completely closed within the cartesian
extrapolation domain and therefore ignores distant connectivities that can contribute
with signal along the line-of-sight. The quiet Sun high altitude corona in which
active regions are embedded is known from off-limb observations \citep[e.g][]{warren2009}
to emit at the 1--2$\times10^6$\,K temperature range. Similarly we expect an excess
335\,\AA\ intensity from small low lying loops that are not part of the extrapolation
for their small footpoint field strength, but contribute to the line-of-sight
observed intensity. One simple way of estimating these contributions is to look outside
active regions. In those areas 335\,\AA\ images have intensities in the range
1.5 -- 5 DN\,s$^{-1}$. Figure~\ref{fig:simF-Lbackg} shows a rendition of the model
where a fixed background of 1.5 DN\,s$^{-1}$  (335 \AA) has been added to every pixel
before noise. This is sufficient to prevent the drop of intensity at low
fluxes where the background has a larger relative impact in each pixel, but it has
no significant effect in the integration at high counts.
The background outside active regions in \ion{Fe}{18} images is at the noise level,
under 1 DN\,s$^{-1}$. While negligible, in Figure~\ref{fig:simF-Lbackg} we show the
effect of a 0.1 DN\,s$^{-1}$ \ion{Fe}{18} background to maintain the symmetry in
the analysis for both filters. It is evident that at low magnetic fluxes,
corresponding to weaker fields, larger polarity separation and therefore smaller
heating as prescribed, the background contribution can even dominate. Note that
for the smallest magnetic fluxes, as the region decays, the \ion{Fe}{18}
counts are dominated by contribution from the wings in the noise distribution leaking
above the threshold. Indeed, the addition of the background accentuates this effect
and raises the counts further introducing new data points in Figure~\ref{fig:simF-Lbackg}
for the final time steps (lower fluxes) of regions 11150, 11259, 11484, 11672, 11765.
Originally those images had count levels under the threshold and therefore were not
included in Figure~\ref{fig:simF-L}.

\begin{figure*}
\centering
\includegraphics[width=\textwidth]{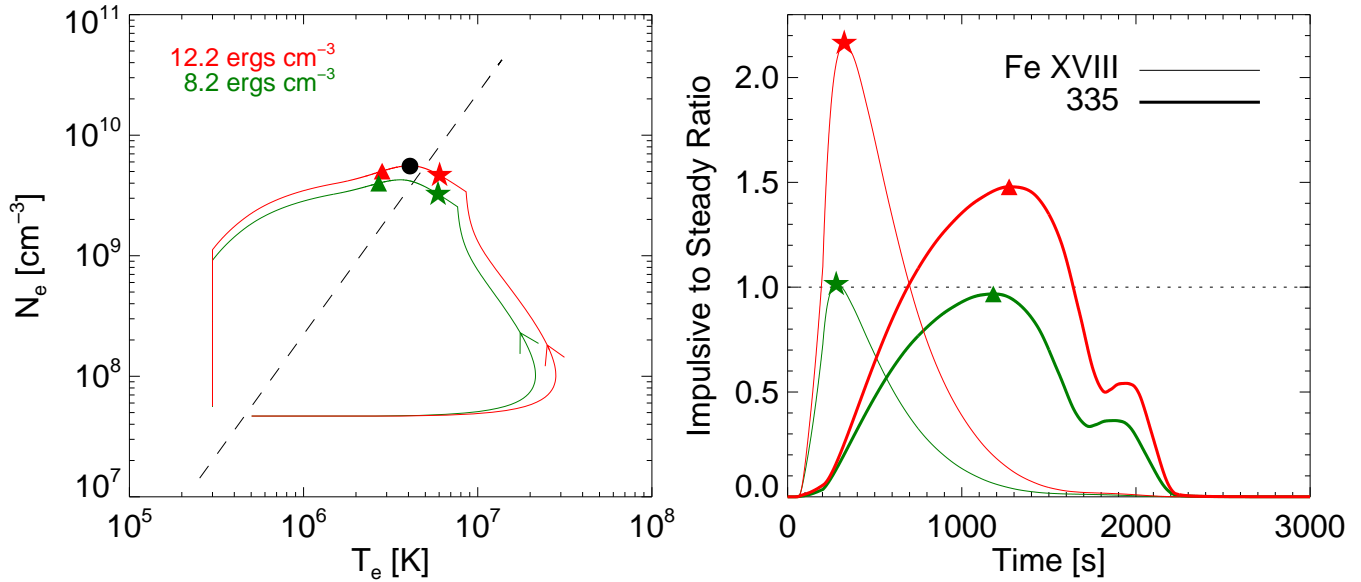}
\caption{EBTEL simulations of a 59 Mm loop heated impulsively. The different colors
represent two different volumetric heating energies. The left panel shows the density
and temperature evolution, where the dashed line illustrates the hydrostatic solutions
and the filled circle is the steady heating in our model. Arrows indicate direction of
evolution. The right panel shows the
ratio of the synthetic lightcurves for \ion{Fe}{18} and 335\,\AA\ with respect to the
steady case (dotted line). The star and triangle symbols are references across the panels.
\label{fig:ebtel-imp}}
\end{figure*}

Assuming that there are no calibration issues that could affect the relative intensities
of \ion{Fe}{18} and 335\,\AA, the systematic underestimation of 335\,\AA\ intensity, a
factor of 3 and 5 in $\Phi/D$ and $\Phi$, is difficult to reconcile with the current model.
Intensities are mainly dependent on density and temperature. In the current set up, to
raise the 335\,\AA\ signal up to the observed values, we can either
increase the density of existing loops, bringing the loop temperatures closer to
the 335 \AA\ filter response and away from \ion{Fe}{18}, or increase the volume
of emission, i.e. the number of loops. Increasing the energy in the steady heating
case has the effect of increasing simultaneously the density and temperature of
the steady state. At the temperatures  of 2 -- 4$\times10^6$\,K that the loops are already
tuned to, this results in higher \ion{Fe}{18} intensities and a competing effect
in  335 \AA\ between the higher number of photons from the density enhancement and
the smaller response of the filter at the growing temperature. The reverse,
decreasing the energy, produces smaller than predicted \ion{Fe}{18} intensities.
Increasing the number of steady loops, by means of filling more coronal volume with them,
would in principle work in favor of increasing 335\,\AA\ signal because the space
to fill grows with height and longer loops are cooler due to the $L^{-1}$ dependence of
the heating. It seems unlikely given that we are missing 2/3 of the signal and
the longer loops are dimmer, but we can not completely rule out that steady heating
could work if we were significantly underestimating the emitting volume.

There is, however, a more attractive explanation of the discrepancy and that is
the choice of the heating model. Our choice of steady heating is based in simplicity
and its past success in reproducing particular features of full active region modeling.
Decades of studies of loop dynamics have, nevertheless, given us confidence that
impulsive heating is likely to play a fundamental role in the evolution of the
EUV loops observed at $\lesssim$2$\times10^6$\,K \citep[see review and references in][]{reale2014}.
Steady footpoint heating has also been proposed to
describe these observations \citep[see recent discussion][]{mikic2013,klimchuk2010}.
The nature of the heating is ultimately likely to be impulsive and any effective
steadiness is just a consequence of the frequency of recurrence of those events
\citep[e.g.][]{klimchuk2006}, with a probable scenario where various degrees of
steadiness coexist and an intermediate case (heating frequency of the order
of the cooling time) dominates \citep{cargill2015}. It is beyond the reach
of this paper to implement impulsive heating, a significantly
larger parameter space. We plan to do it in a
forthcoming paper. We can show, however, that the effect of the impulsiveness
would in fact be to move the intensities in the direction for a better match.

Figure~\ref{fig:ebtel-imp} shows the EBTEL calculations for a 59 Mm loop taken from one
of the initial snapshots in NOAA 11726. The filled circle represents the density
and temperature solution in the steady case, as simulated throughout this paper. That
density-temperature pair falls very close to the density
and temperature solutions (dashed line) of hydrostatic equilibrium
\citep{rosner1978}. In the impulsive case, density and temperature change as a
function of time (red and green lines for two energy values) and as pointed out by
\citet{winebarger2004} the loop remains underdense before reaching its equilibrium
point, at $\sim$4$\times10^6$\,K in this case.
At that temperature and density, when the radiative losses begin to dominate, it is
when the loop will radiate more in 335\,\AA, that is $\sim$700s into the time evolution of
the red curve in Figure~\ref{fig:ebtel-imp}. The loop reaches the formation temperature
of \ion{Fe}{18} (7$\times10^6$\,K) earlier, within the underdense section of the evolution.
Therefore, one could in principle conclude that with an impulsive heating model, the
loop will be underdense when emitting \ion{Fe}{18} and the adjustment in energy needed
to match these new set of intensities would decrease the relative difference to 335\,\AA.
This turns out to be true, but the explanation is not just in the density of the loop
at the different stages. In the right panel of Figure~\ref{fig:ebtel-imp} we show the ratio
of \ion{Fe}{18} and 335\,\AA\ intensities to the steady case. If we fix the energy
to match the peak \ion{Fe}{18} with the steady case (green curves) we see that both
filters produce a ratio of 1 at peak, apparently disproving the differential effect.
In the impulsive case, it is the combined effect of density and temperature sensitivity
of the filter, where a differential factor is introduced, a factor of 2 in this
particular example. The loops radiate longer in 335\,\AA. It is the time integrated
intensity that raises the 335\,\AA\ emission with respect to \ion{Fe}{18}. In other words,
with impulsive heating at any given time and line-of-sight we would expect to see
more 335\,\AA\ loops, a result that works in the direction of reconciling the
discrepancy that we find in the steady case. This needs to be demonstrated in a
full, time-dependent active region simulation. The exercise in any case shows that
\ion{Fe}{18} and \ion{Fe}{16} are diagnostics of different phases in the loop evolution.

\section{Conclusions}
We present new measurements of the magnetic flux - EUV radiance relationship
of solar active regions from observations of the AIA and HMI instruments on the
SDO mission in the 335\,\AA\ passband and the \ion{Fe}{18} component of the
94\,\AA\ band. Nine active regions were observed over
several stages of their evolution from birth to decay. We confirm past reports that
a power law ($I\propto\Phi^{\alpha}$) is a good representation of the correlation and
extend them to the \ion{Fe}{18} spectral line, a particularly good diagnostic for
coronal heating in active regions, as it forms at a temperature very close to the
characteristic peak of the emission measure distribution. We find, in fact, that
a better indicator of the \ion{Fe}{18} radiance is the ratio of the total unsigned
magnetic flux to the polarities separation ($\Phi/D$) with a slope of $\alpha=3.22\pm0.03$.

We then use these results to test our current understanding of active region
magnetic field evolution and coronal heating. We use magnetograms, from simulations
of the magnetic flux decay of these nine active regions produced by the Advective
Flux Transport model in \citetalias{ugarte-urra2015}, as a boundary condition for
the potential magnetic topology in the coronal atmosphere. We then model
the hydrodynamics of each field line independently with the EBTEL 0D model assuming
steady heating that scales as $\bar{B}/L$. Finally, we integrate all loops to calculate
the active regions' emission in \ion{Fe}{18} and 335\AA\ and compare them to the
magnetic fluxes and polarities separation from the AFT simulated magnetograms.
We find that steady heating is able to partially reproduce slopes of the flux - radiance
relationship for both lines, but find discrepancies on the magnitudes that we
speculate could be resolved with impulsive heating, although this needs to
be demonstrated in the future.

These results support the idea that our understanding of fundamental processes
in the Sun such as the transport of magnetic fields on the solar surface, the
distribution of magnetic fields in the corona and the plasma hydrodynamics along
the loops have reached a level of maturity that allows us to couple them and do
detailed quantitative comparisons to available data. There is significant progress
to be made in all three aspects of the model, particularly in the topology and heating
parts of our experiment that can already be upgraded with available tools such as
NLFF and time dependent heating. This will be part of our future active region
studies. Ultimately, we expect that 3D MHD models based on first principles will
reach a point where they can address the spatial and temporal scales necessary to
be tested in a similar manner as our quasi-static steady heating set up. Spectroscopy
of high temperature lines such as \ion{Fe}{18} is a promising diagnostic of heating
to confront to those models.


\acknowledgments
We would like to thank the anonymous referee for all the comments and suggestions that
helped improve the paper.
AIA and HMI data are courtesy of NASA/SDO and the AIA and HMI science teams.
I.U.U. was supported by NASA Heliophysics Guest Investigator program. L.A.U. was supported
by the NASA Heliophysics Guest Investigator program and by the National Science Foundation
Atmospheric and Geospace Sciences Postdoctoral Research Fellowship Program. The National
Center for Atmospheric Research is sponsored by the National Science Foundation. P.R.Y. was
supported by NASA grant NNX13AE06G.

\vspace{5mm}
\facilities{SDO (AIA,HMI)}
\software{CHIANTI \citep{dere1997}, SolarSoft \citep{freeland2012}}

\bibliography{main.bib}
\bibliographystyle{apj}

\end{document}